\documentclass[%
 reprint,
superscriptaddress,
 amsmath,amssymb,
 aps,
]{revtex4-2}

\usepackage{graphicx}
\usepackage{dcolumn}
\usepackage{bm}

\usepackage[dvipsnames]{xcolor}
\usepackage{stackrel}

\usepackage{amsfonts}

\usepackage[normalem]{ulem} 

\usepackage[version=3]{mhchem}

\newcommand{\TSub}{\textsubscript}

\newcommand{\dpht}{$\Delta\mathrm{pH}$}
\newcommand{\dph}{\Delta\mathrm{pH}}
\newcommand{\Vpp}{{V\TSub{pp}}}
\newcommand{\wvp}{\% \mathrm{~w/v}}

\usepackage{xcolor}

\usepackage[caption=false]{subfig}

\begin{document}

\preprint{APS/123-QED}

\title{Artificial Chemotaxis under Electrodiffusiophoresis}





\author{Carlos A. Silvera Batista}
\affiliation{Department of Chemical and Biomolecular Engineering, Vanderbilt University, Nashville, Tennessee 37212, United States.}
\affiliation{Vanderbilt Institute for Nanoscale Science and Engineering, Vanderbilt University, Nashville, Tennessee 37212, United States}
\email{silvera.batista@vanderbilt.edu}

\author{Kun Wang}
\affiliation{Department of Chemical and Biomolecular Engineering, Vanderbilt University, Nashville, Tennessee 37212, United States.}

\author{Hannah Blake}
\affiliation{Department of Chemical and Biomolecular Engineering, Vanderbilt University, Nashville, Tennessee 37212, United States.}

\author{Vivian Nwosu-Madueke}
\affiliation{Department of Chemical and Biomolecular Engineering, Vanderbilt University, Nashville, Tennessee 37212, United States.}

\author{Sophie Marbach}
\affiliation{CNRS, Sorbonne Universit\'{e}, Physicochimie des Electrolytes et Nanosyst\`{e}mes Interfaciaux, F-75005 Paris, France}
\email{sophie.marbach@cnrs.fr}

\date{\today}

\begin{abstract}
Diffusiophoretic motion induced by gradients of dissolved species has enabled the manipulation of colloids over large distances, spanning hundreds of microns. Nonetheless, studies have primarily focused on simple geometries that feature 1D gradients of solutes generated by reactions or selective dissolution. Thus, our understanding of 3D diffusiophoresis remains elusive despite its importance 
in wide-ranging scenarios, such as cellular transport and nano-fluidics. Herein, we present a strategy to generate 3D chemical gradients under electric fields.  In this approach, faradaic reactions at electrodes induce \textit{global} pH gradients that drive long-range transport through electrodiffusiophoresis. Simultaneously, the electric field induces \textit{local} pH gradients by driving the particle's double layer far from equilibrium. As a result, while global pH gradients lead to 2D focusing away from electrodes, local pH gradients induce aggregation in the third dimension. Resulting interparticle interactions display a strong dependence on surface chemistry, and particle size. Furthermore, pH gradients can be readily tuned by adjusting the voltage and frequency of the electric field. For large P\'eclet numbers, we observed a chemotactic-like collapse. Remarkably, such collapse occurs without reactions at a particle's surface. By mixing particles with different sizes, we also demonstrate the emergence of non-reciprocal interactions through experiments and Brownian dynamics simulations. These findings suggest a wide array of possibilities for the dynamic assembly of materials and the design of responsive matter. 
\end{abstract}

\maketitle

The migration of cells along chemical gradients, \textit{chemotaxis}, constitutes an essential mechanism of adaptation, that serves to mount an immune response or to sustain embryonic development \cite{Wadhwa.2022,Budrene.1991}.  To navigate a chemical landscape, cells rely on the active generation, detection, transmission and transduction of chemical signals~\cite{SenGupta.2021}. 
Interestingly, the same principles can be replicated through purely physical mechanisms, such as diffusiophoresis (DP) \cite{Anderson:1989gw}. DP refers to the directed, yet passive, transport of colloidal particles in response to concentration gradients of solutes \cite{Derjaguin.1961,Anderson.1982gn,Prieve.1984v9m,marbach2019osmosis,shim2022diffusiophoresis}. Particles ``sense'' solute gradients through surface-solute interactions, which build up fluid pressure and induce flows. 

DP constitutes a versatile form of colloidal transport because of the possibility of controlling motion by tuning the properties of particles and solute gradients. Consequently, interest in DP has increased over the last 15 years for applications in separation \cite{Shin:2017drd,Florea:2014ega}, transport in porous media \cite{Kar:2015gn,Shin:2016ee}, detergency \cite{Shin:2018il,Warren:2018cj}, drug delivery \cite{Joseph:2017iv}, desalination~\cite{Guha:201524d,prieve2019diffusiophoresis}, decontamination~\cite{fu2021electrokinetic}, and patterning \cite{Alessio.2023}.
Fascinatingly, DP is also harnessed by cells to transport cargos~\cite{ramm2021diffusiophoretic} or to navigate chemical gradients~\cite{doan2020trace}, undergoing sometimes both DP and chemotaxis~\cite{shim2022diffusiophoresis}.

External and global chemical gradients can be generated through many mechanisms \cite{Velegol:2016et}, such as chemical reactions \cite{Nizkaya.2022, Brooks.2019}, mineral dissolution \cite{McDermott.2012}, and molecular exclusion \cite{Florea:2014ega,Niu:2018gp}. Lately, solute-inertial beacons, which release solutes continuously, have facilitated the study of long-range DP~\cite{Banerjee:2016ce,Banerjee.2019hn,Niu:2018gp,Niu:2017ff}.  Solute-inertial beacons impose deterministic—attractive or repulsive—forces over long distances, $\gg 1~\mu$m \cite{Niu:2017ff,Banerjee.2019z1c}. In contrast, local chemical fields can arise when microparticles act as  sinks or sources of solutes. For example, photocatalytic reactions on the surface of particles promote on-demand colloidal phase separation \cite{Mu.2022,massana2018active}, with applications in optical camouflage~\cite{Zheng.2023}. For particles with broken symmetry, anisotropic consumption of solutes induces phoretic self-propulsion~\cite{Moran.2017}, and rich collective dynamics, exemplified by motility-induced phase separation, chemotactic collapse or gel-like phases~\cite{Fadda.2023,theurkauff2012dynamic,Pohl:2014hn}. However, all the above systems  require the use of a solute playing the role of a slowly consumed fuel; therefore, solute gradients are hard to sustain for long periods of time.

Despite existing diverse strategies to induce local and global chemical gradients, studies on DP mostly focus on solute gradients that are either in simple 1D geometries, or that emanate radially from a beacon or particle. Nonetheless, recent theoretical studies reveal that unconventional geometries open new opportunities for complex manipulation of colloids. Warren predicted 2D or 3D orthogonal gradients of two salts can induce solenoidal currents~\cite{Warren.2020}, an effect which was recently demonstrated experimentally~\cite{williams2024colloidal}. Calculations by Raj et al. showed that judicious placement of sources and sinks enables the spatial and temporal design of  2D diffusiophoretic banding~\cite{Raj.2023}. Therefore, 3D diffusiophoresis can extend the range of chemically-engineered colloidal manipulation.

\begin{figure*}[htp!]
\includegraphics[trim=0 0 0 0,clip,width=17.5 cm]{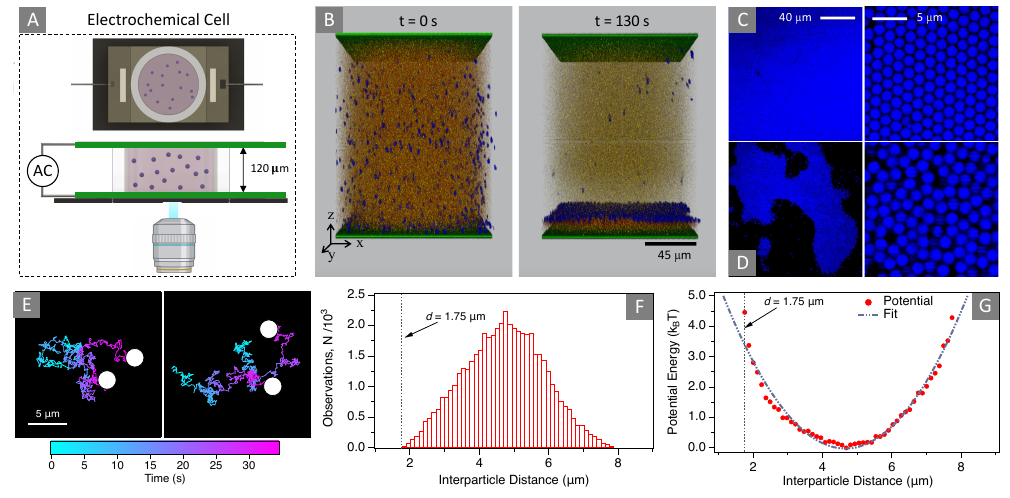}
 \centering
\caption{\label{Fig:1} \footnotesize \textbf{EDP induces strong in-plane pairwise colloidal attraction.} (A) Schematic of experimental setup. (B) Confocal images of volume between the electrodes (green slabs) before and 130 s after applying an AC field (100 Hz, 5 V\TSub{pp}). CB-PS particles (blue) focus on a layer located tens of microns from the bottom electrode, where pH changes most rapidly. (C) At high number densities, particles form crystals in the levitated layer. (D) Even at lower concentrations, particles form aggregates with low crystalline order. (E) Trajectories of dimers in the levitated layer. (F) Histogram of interparticle distance for dimers, averaged over 4 different pairs totaling 50000 frames. Panel (G) shows the pairwise potential energy of dimers obtained by inverting the Boltzmann distribution. The fitted curve corresponds to $\phi(r) = k (r-r_{\rm min})^2/2$, here $k = 3.3~\mathrm{nN/m}$ and $r_{\rm min} = 4.6~\mathrm{\mu m}$. Data in panels E-G was acquired from experiments under dilute conditions, using 1.75 $\mu$m CB-PS particles, 100 Hz and 5 \Vpp. The concentration of particles was $5\times10^{-4}\wvp$.}
\end{figure*}

In this work, we introduce a strategy to induce 3D gradients of solutes that facilitate  versatile pairwise interactions and assembly.  
The strategy, electrodiffusiophoresis (EDP), combines electrophoresis (motion under an applied electric field) and diffusiophoresis \cite{Dukhin:1980,Dukhin:1982,Dukhin:1982b,Ulberg.1990,Rica:2010jz,Tricoli:2015hv,fu2021electrokinetic}.
Using low-frequency AC fields ($\lesssim$ 2 kHz), we induce \textit{global} pH gradients that lead to long-range transport and focusing of charged particles on a plane~\cite{Wang.2022,Wang.2022p2}. Once at the focusing plane, the presence of particles perturbs \textit{local} pH gradients, thus
resulting in long-range in-plane attraction and ``artificial'' chemotaxis. Remarkably, we induce local chemical fields without  consuming a fuel on the surface of particles. Instead, pH gradients result from driving the double layer and its surroundings far from equilibrium. Local pH gradients depend on properties of particles  (surface chemistry, zeta potential, and diameter), and can be easily tuned via the frequency and amplitude of the applied electric field. Under some conditions, we observe collective behavior reminiscent of chemotactic collapse and non-reciprocal interactions, indicating a broad set of possibilities to unlock and engineer diffusiophoresis by externally-controlled 3D chemical profiles.



\subsection*{Soft and long-range interparticle interactions under EDP}

~Herein, carboxyl-polystyrene particles (CB-PS) are subjected to low-frequency AC fields in parallel plate devices, see Fig.~\ref{Fig:1}A. Figure~\ref{Fig:1}B shows 3D microscopy images of the electrochemical cell before and 130 s after applying an AC field of 5 V\TSub{pp} and 100 Hz, 
viewed from the $x-z$ plane. 
The images combine emission intensity from a ratiometric dye (SNARF-1) collected at 580 nm (I\TSub{1}, yellow orange), and at 640 nm (I\TSub{2}, red), emission intensity from CB-PS particles (blue), as well as reflection from electrodes (green slabs). Once an AC field is applied, changes in color from SNARF-1 indicate substantial changes of pH throughout the electrochemical cell (see SI Sec.~1 and Ref. \cite{Wang.2022p2}). In response to non-monotonic pH profiles, particles migrate about 15 $\mu$m from the bottom electrode, where a maximum pH occurs~\cite{Wang.2022,Wang.2022p2}. Since particles experience vertical potential wells in the order of $100~k_{B}T$, they remain within this levitated layer as long as the electric field is on. If loaded at sufficiently high particle densities ($2.5\wvp$), particles experience a transition from a disordered liquid into a colloidal crystal, as reported in~\cite{Wang.2022p2} (Fig. \ref{Fig:1}C). Surprisingly, even under dilute conditions ($5\times10^{-5}\wvp$), particles form aggregates with high local order (Fig.~\ref{Fig:1}D). This observation reveals that particles not only experience a large $z$-focusing potential, but also pairwise interactions that induce aggregation under dilute conditions.

To reveal this interaction potential, we performed experiments in dilute conditions ($5\times10^{-4}\wvp$) to avoid large aggregates. A pair of nearby interacting particles (a dimer) was first identified in the levitated layer (see SI-Movie 1). The position and trajectories of each particle were then extracted using particle tracking (Fig.~\ref{Fig:1}E). Histograms of interparticle distance reveal that, through Brownian motion, particles sample a range of interparticle distances equivalent to 4 particle diameters ($d = 1.75~\mathrm{\mu m}$), around a typical distance of $r_{\rm min} \approx 4.6~\mathrm{\mu m}$ (Fig. \ref{Fig:1}F, and Fig.~S3). The positional information can be converted in this equilibrium regime into a potential energy landscape, $\phi(r)$, experienced by the particles, using the Boltzmann equation: 
\begin{equation}
 P\left(r\right)=Ae^{\phi(r)/(k_B T)}
  \label{eqn:PotentialDist},
\end{equation}
where $P(r)$ is the probability of sampling interparticle positions at a distance $r$, whereas $A$ is a normalization constant such that $\int_{0}^{\infty}{P(r)dr=1}$, see also SI Sec.~2 for details. Figure \ref{Fig:1}G shows the potential energy landscape calculated from the distribution in Fig.~\ref{Fig:1}F \cite{JeffreyAFagan:2002bk,Rupp:2018kl}. Particles experience an attractive confining potential, with a depth of a few $k_{B}T$. This interaction has two important features. First, it is long-range (a few microns) compared to either electrostatic or van der Waals colloidal interactions (which extend over a few nanometers for micron-sized particles)~\cite{derjaguin1941,verwey1948theory,parsegian2005van}. It is important to note that particles experience an even larger range of interactions than we can resolve. In fact, due to the depth of the potential, events where particles move far from each other are rare and hard to sample. Second, the potential is soft, with a typical curvature corresponding to a spring force $k \simeq 3-4 \mathrm{~nN/m}$ (dictating an interparticle potential $\phi(r) = k (r-r_{\rm min})^2/2$), orders of magnitude smaller than a polymer spring constant~\cite{rubinsten2003polymer,cui2022comprehensive}. Interaction potentials of a few $k_{B}T$ are conducive to achieving well-organized structures through directed assembly~\cite{Bevan.2011}. Next, our goal is to understand the fundamental origin of this peculiar soft and long-range interaction potential. 

\begin{figure*}[htp!] 
\includegraphics[trim=0 0 0 0,clip,width=0.9\textwidth]{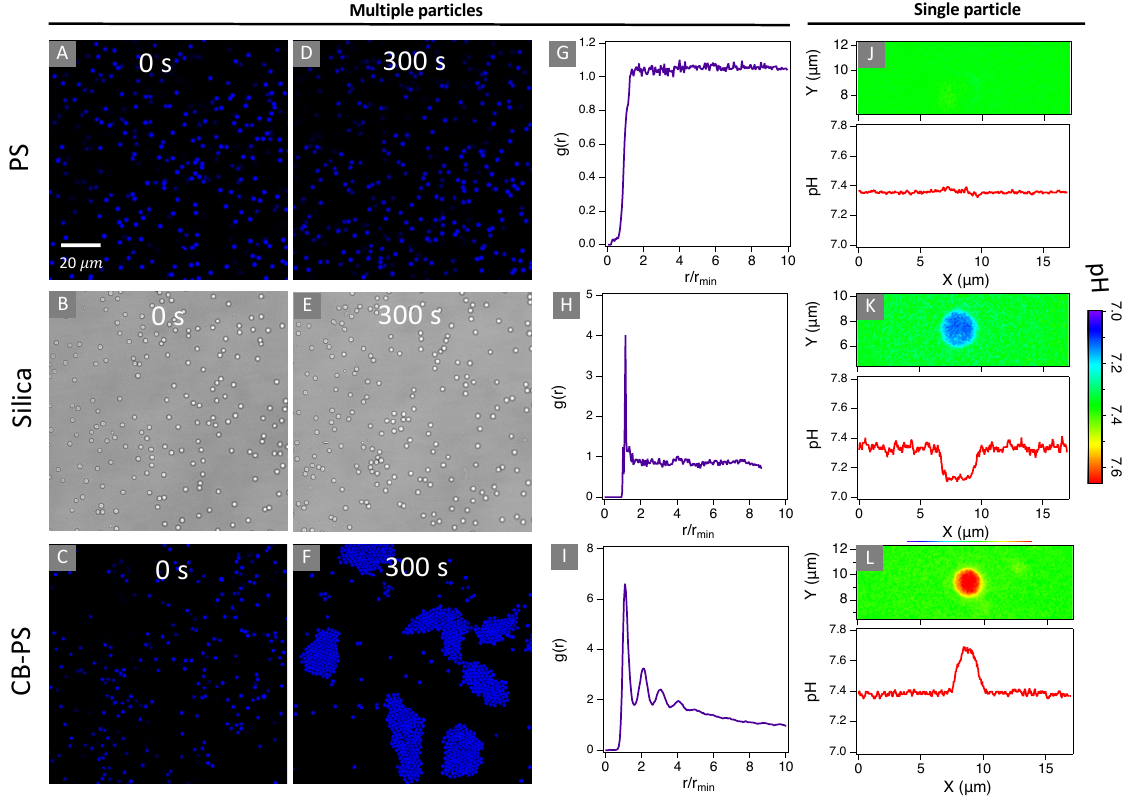}
 \centering
\caption{\label{Fig:2}\footnotesize \textbf{Multiscale experiments reveal surface chemistry impacts the resulting interparticle potential under EDP.} (A-C) Fluorescense and brightfield images of polystyrene (PS), Silica, and carboxyl-functionalized PS particles immediately after reaching the levitated layer and (D-F) 300 s after. (G-I) Pairwise radial distribution function, $g(r)$, for each type of particles. (J-L) 2D heat maps indicating no change, decrease, or increase of pH around particles. Particles have similar sizes, zeta potentials and dielectric constants. A key difference between particles is the pK\TSub{a} of dominant surface groups; for sulfate, silanol and carboxyl groups, the approximate values of pK\TSub{a} are -1, 2.5, and 4. The applied field was 100 Hz and 5 \Vpp. The $g(r)$ was obtained from averages over 165 frames, obtained after 300 s, for samples with concentrations of approximately $5\times10^{-3}\wvp$. Fluorescense images for PS and CB-PS particles are in false colors.   
}
\end{figure*}

\subsection*{Interactions originate from surface chemistry, through the establishment of a local pH gradient}

To determine the origin of the interparticle potential, we repeat the experiment with different particles in moderately dilute conditions ($\approx 5\times10^{-3}\wvp$). The model systems are plain polysyrene (PS), silica and carboxyl-functionalized PS (CB-PS) particles. Images in Fig.~\ref{Fig:2}A-F compare  distributions of particles as soon as they arrive in the levitated layer and 300 s after. In stark contrast to CB-PS particles, neither plain PS nor silica particles form readily discernible structures over the course of the experiment.  To gain further insight, we calculate, in the equilibrated state after 300 s, the radial pair distribution function $g(r)$,  which quantifies the probability of finding a particle at a distance, $r$, from a reference particle (Fig.~\ref{Fig:2}G-I).  For plain PS, $g(r)$ resembles the behavior of an ideal hard-sphere gas since it remains constant with a value near 1; that is, positions of particles  remain uncorrelated. For silica, $g(r)$ displays a sharp, rapidly decaying peak, reminiscent of a fluid-like structure with weak interactions. The positions of silica particles, albeit more correlated than for plain PS---as can be seen through small clusters in Fig.~\ref{Fig:2}E---display only short ranged correlations. CB-PS particles display regular peaks, decaying in magnitude at larger distances, corresponding to ordering in dense fluid suspensions~\cite{Hansen.2013}.  

Since $g(r)$ is a manifestation of interparticle potentials~\cite{Attard.2002,Hansen.2013}, we can infer that attractive pairwise interactions are much smaller in range and magnitude for silica than for CB-PS, and nearly nonexistent for PS. Since all these particles have comparable size, surface charge and dielectric constants (see Methods), the differences in pair potentials must originate from another source. The surface chemistry of particles constitutes a plausible source since the dominant charged surface groups for PS, silica and CB-PS particles are ester sulfates, silanols and carboxylates, with dissociation constants (pKa) approximately equal to -1, 2.5, and 4, respectively.  

To gain further insight into the effect of surface chemistry, we diluted suspensions to a concentration of  $5.0\times{10}^{-4}$ to examine individual levitating particles. Fluorescence intensity from SNARF-1 surrounding each particle was then converted to pH values using ratiometric analysis, see Methods. The heat maps in Fig.~\ref{Fig:2}J-L capture the spatial distribution of pH around a single particle; blue and red colors represent pH extrema corresponding to 7.0 and 7.6 respectively. Strikingly, a significant change of pH occurs near silica and CB-PS particles, while there is no observable change in pH near PS particles. The change in pH for silica particles is small, $\simeq 0.1$ units, and negative compared to that for CB-PS particles, which is large, $\simeq 0.3$ units, and positive. A correlation thus appears to exist between the observed changes in pH and $g(r)$: CB-PS particles display the largest change in pH and a strong tendency to aggregate, whereas PS particles show minimal change in pH, with no aggregation. Meanwhile, silica particles display intermediate behavior. Consequently, these observations suggest local pH gradients underlie the observed interaction potential.

\subsection*{Local pH gradients mediate long-range interactions via diffusiophoresis}

We hypothesize the interaction potential stems from diffusiophoretic flows, induced by the local pH gradients surrounding the particles, in which each particle acts as a beacon. 
In our previous publication, we demonstrated that pH gradients along the z-axis cause particle migration and focusing via diffusiophoresis (DP)~\cite{Wang.2022p2}. Similarly, a target particle should respond to local chemical gradients induced by a neighboring particle; that is, particles should respond to multidimensional pH gradients as well.   
Generally, the diffusiophoretic velocity ($U_{\rm DP}$) of a particle in a uniform and constant concentration gradient is given by,
\begin{equation}
   U_{\mathrm{DP}}=\mu_{\rm \rm DP}\nabla \ln c,
\end{equation}
where the magnitude and sign of the diffusiophoretic mobility ($\mu_{\rm DP}$) depend on surface chemistry of particles, as well as on concentration ($c$) and properties of solutes~\cite{marbach2019osmosis}. 

Here, the measured pH gradient corresponds to a concentration gradient of \ce{H+}. Since the in-plane pH gradient is symmetric around the particle, $U_{\mathrm{DP}}$ is also symmetric and depends only on interparticle distance $r$. 

\begin{figure*}[htp!]
\includegraphics[trim=10 25 0 0,clip,width=18cm]{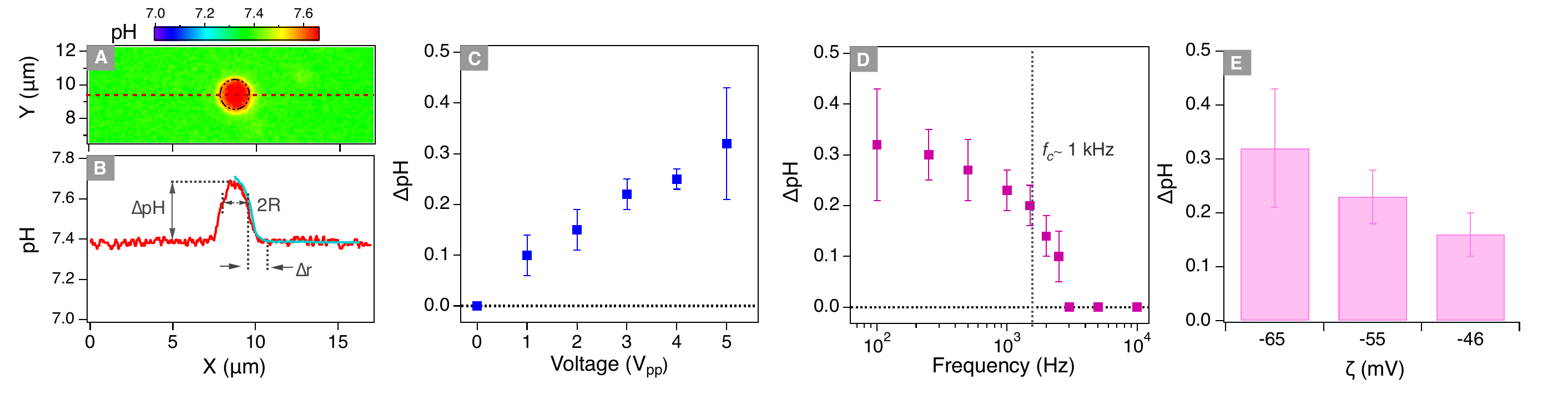}
 \centering
\caption{\label{Fig:3}\footnotesize \textbf{Parameters of AC field  and surface charge determine local pH gradients.} (A-B) The same pH map and 1D pH profiles as in Fig.2L; notice the outline of the particle in Panel A via a dashed line.  The dashed horizontal line in Panel B provides the visual definition of $\dph$ in Panel A. The cyan curve represents the radial pH profiles averaged over all polar angles (see SI, Fig. S5). Experimental conditions were as follows: C) 100 Hz, D) 5 ~\Vpp, and E) 100 Hz, 5 \Vpp. Error bars in Panels C-E represent the standard deviation of values from ten frames and ten different particles, for a total of 100 points. Zeta potential of particles was tuned by cross-linking PEG chains of different molecular weights to the carboxyl groups. Experiments in Panels C-D were performed using 5 $\mu$m CB-PS particles, for samples with concentrations of approximately $5\times10^{-4}\wvp$.}
\end{figure*}

We can estimate $\nabla \ln c \simeq \dph / \Delta r$ by taking $\dph \simeq 0.1$  and $\Delta r \simeq 1~\mathrm{\mu m}$ from experiments (Fig.\ref{Fig:3}A-B). For our  particles,  $\mu_{\rm DP}$ typically ranges from 10 to 100 $\mathrm{~\mu m^2/s}$~\cite{marbach2019osmosis}. Using these values, we estimate  $U_{\rm DP}(R) \simeq 1-10~\mathrm{\mu m/s}$ at the particle's surface, which agrees with the range of measured interparticle velocities (Fig. S3).
 Because fluid flow is incompressible, radially symmetric diffusiophoretic velocities should scale as $U_{\rm DP}(r) \approx U_{\rm DP}(R) \, R^2/r^2$~\cite{niu2017microfluidic,Niu:2017ff,Niu:2018gp}. Such a diffusiophoretic flow field can be translated via $U_{\rm DP}(r) \approx - D_0 \phi'(r)/k_B T$ into a long-range effective potential $\phi(r) \sim 1/r$, where $D_0$ represents the particle's diffusion coefficient. Note that diffusiophoresis is a force-free mechanism; therefore, the potential is merely an effective representation of the underlying flows~\cite{Niu:2017ff,marbach2019osmosis}. Considering these scaling arguments, it is reasonable to conclude that local pH gradients induce  aggregation of particles via diffusiophoresis.

\subsection*{External AC forcing sets local pH gradients around charged particles}

We now investigate the origin of the local pH gradient surrounding CB-PS particles, which exhibit the strongest response to AC fields. To quantify the pH gradient, we extract pH profiles across the particle's equator (Fig.~\ref{Fig:3}A-B). The disturbance of pH extends over 1.0 $\mu$m ($\Delta r$) from the surface of particles, making it of a similar length scale as the measured pairwise interaction potential. The profiles also provide \dpht, the difference between the maximum pH value at the particle surface and bulk pH.

The properties of the external AC field---amplitude and frequency---set the values of \dpht. Fig.~\ref{Fig:3}C shows \dpht~for CB-PS particles under different voltages. First, it is apparent that pH gradients do not appear in the absence of an applied electric field. Second, for voltages as low as 1 V\TSub{pp}, pH changes substantially near a particle's surface, and increases with increasing voltage. Experiments at different frequencies (Fig. \ref{Fig:3}D) show pH gradients remain relatively constant between 0.1 and 1 kHz before rapidly decreasing at a cutoff frequency, $f_c \simeq 1~\mathrm{kHz}$. This relaxation occurs at a similar frequency range for 1.75 $\mu$m (see SI, Fig. S10).  The range of the variation surrounding the particle ($\Delta r$) is much less sensitive to AC field parameters. Therefore, there is a wide frequency window to manipulate this phenomenon. 

To discern the role of surface charge, we modify particles with poly(ethylene glycol) (PEG). The PEG chains controllably modify the amount of charges by covalently attaching to carboxyl groups. Consequently, the zeta potential of particles decreases as the size of attached PEG molecules increases. Figure \ref{Fig:3}E indicates \dpht~increases with increasing magnitude of zeta potential. Local pH gradients thus depend on the number of carboxyl groups available.

\begin{figure*}[htp!]
\includegraphics[trim=0 0 0 0,clip,width=\textwidth]{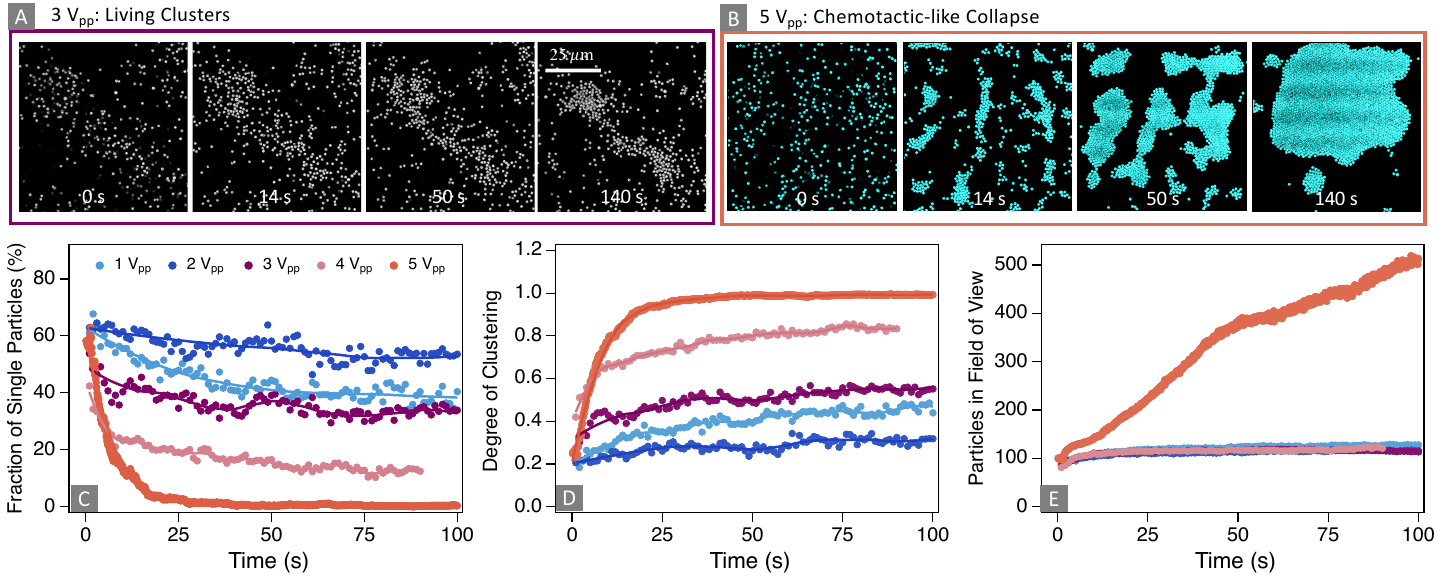}
 \centering
\caption{\label{Fig:4}\footnotesize \textbf{Versatile nonequilibirum aggregation: from living clusters to chemotactic collapse.} Sequence of images for 1.75 $\mu$m CB-PS particles, and 100 Hz: (A) at 3 \Vpp~ demonstrating transient aggregation, and (B) at 5 \Vpp~demonstrating rapid coarsening. Characteristic horizontal black-magenta shades at 140~s correspond to waves on the assembled 2D micro sheet, with a wavelength around $\lambda \simeq 25~\mathrm{\mu m}$. These waves represent slight variations in focusing position of particles. (C) Fraction of single particles defined as the number of single particles relative to the total number of particles in the field of view at a given time. (D) Degree of clustering, as introduced in Ref.~\cite{mognetti2013living}, $\theta(t) = 1 - 1/\langle n_{\rm clust}(t)\rangle $ where $\langle n_{\rm clust}(t)\rangle$ is the mean cluster size at a given time. (E) Total number of particles in the field of view, rescaled by the number of particles in the field of view at the beginning of image acquisition.  In (C-E), cluster properties are measured with increasing \Vpp, at 100 Hz, with the legend shared across plots. Dots are data points and lines are guides for the eye corresponding to window averages. These experiments were performed with 1.75 $\mu$m CB-PS particles, with concentration of approximately  $5\times10^{-2}\wvp$, and an average $\zeta = -65.0 \pm 3.1$ mV. For clarity, fluorescence images for 3 and 5 \Vpp~ are reported in grayscale and false colors, respectively.}
\end{figure*}

These observations provide important insights. First, pH gradients only arise out of equilibrium, where the energy source corresponds to the finite electric field. Second, changes of \dpht~ with zeta potential confirm the importance of available carboxyl groups. Therefore, we surmise charge regulation and dissociation equilibria of surface groups play important roles. Third, changes of \dpht~ with frequency provide the characteristic time scale for the underlying phenomenon.

A diffusive process that matches the measured time scales is the relaxation of a polarized electrical double layer. When the double layer near a highly charged particle is driven far from equilibrium by an electric field, concentration gradients of electrolyte are established to compensate for the difference in conductivity near and far from the surface of the particles; this phenomenon is called concentration polarization, CP. The concentration gradients of the neutral electrolyte extend beyond the double layer, have characteristic lengths in the order of particle’s diameter, and their relaxation is dictated by diffusive timescales~\cite{kamsma2023iontronic,Dukhin.1993,Lyklema:1995v2}.  Therefore, the characteristic time ($\tau$) for CP is dictated by the diffusion of ions across a distance roughly equal to the particle’s diameter, $d^2/D$. Taking $d \simeq 2 \mathrm{~\mu m}$, and the diffusion coefficient for \ce{H+}, $D \simeq 9 \times 10^{-9}~\mathrm{m^2/s}$,  the characteristic time is thus $\tau=d^2/D\simeq 0.4 \mathrm{~ms}$, amounting to a frequency of $1/\tau \simeq 2.5~\mathrm{kHz}$. Remarkably, this frequency agrees with the cut-off values observed in Fig.\ref{Fig:3}D and Fig. S10. The pH gradients observed near particles are likely a manifestation of the long-range concentration fields induced by CP. Furthermore, the increase of \dpht~ with magnitude of zeta potential follows the expectation that a higher surface charge---higher surface conduction---leads to larger gradients. Consequently, we conjecture the observed \dpht~ results from driving the double layer of dissociating groups far from equilibrium under highly non-monotonic global pH profiles.

\subsection*{Manipulating aggregation under EDP: from living clusters to chemotactic-like collapse}

Linking changes in pH to interaction potential widens the range and versatility of field-driven interactions. Images in Figure \ref{Fig:4}A and B contrast the response of CB-PS particles (1.75 $\mu$m) under 100 Hz and different voltages.  At 3 \Vpp~ and moderate concentrations ($5\times10^{-2}\wvp$), particles rapidly form aggregates (SI-Movie 2). Visually, the average aggregate size ceases to grow after some time. Particles in aggregates remain highly dynamic, with some  coming in and out frequently. The aggregates resemble the behavior of the so-called ``living clusters'' seen in systems of self-propelled asymmetric particles~\cite{Pohl:2014hn,mognetti2013living,buttinoni2013dynamical,redner2013reentrant,stenhammar2013continuum,fily2012athermal,stark2018artificial,theurkauff2012dynamic}.

In constrast, at large voltages (5 \Vpp) the system does not reach a stationary state.
After a few seconds, small, tight, clusters begin to form. These small clusters then attract more individual particles (14 s, Fig.~\ref{Fig:4}B). The small clusters then aggregate with one another (50 s), in a similar way as the coarsening of droplets under Oswald ripening, eventually forming a single big cluster (140 s). Astoundingly, this cluster is not in a stationary state, as it keeps on attracting smaller clusters beyond the field of view (SI-Movie 3). 

This tendency to collapse is reminiscent of the so-called ``chemotactic collapse'' displayed by bacteria or chemically-fueled, active Brownian particles (ABPs). Simulations of ABPs with local chemical fields have predicted collapsing behavior for small P\'eclet numbers, that is, when the propulsion velocity is slightly larger than the ``Brownian velocity''~\cite{Fadda.2023,Pohl:2014hn}. Here, the out-of-equilibrium aggregation phenomenon is radically different from ABPs. Following the discussion in Ref.~\cite{mognetti2013living}, while \textit{activity tends to break apart ABP clusters} via increased self-propulsion, \textit{here activity is the driving force for aggregation} and thermal diffusion tends to break clusters. The P\'eclet number in our setting is $\mathrm{Pe} = U_{\rm DP} R / D_0$ and must be large enough to observe collapse. With $U_{\rm DP} \simeq 1 - 10 \mathrm{~\mu m/s}$, we find $\mathrm{Pe} \simeq 10 - 100$ which hints that activated aggregation can indeed be high enough to induce collapse in our experiments. 

 Paradoxically, while chemotactic behavior usually requires fuel consumption, here, there is no fuel! The driving force originates from the electric field that drives particle motion far from equilibrium.  Measurements of pH reveal that \dpht~ increases with the number of particles in a cluster, with a linear increase for clusters of up to $n_{\rm clust} = 10$ particles (SI Fig.~S8). Therefore, the driving force for aggregation increases with cluster size, thus facilitating the migration of smaller clusters to larger ones. This increase in driving force can be recovered by considering long-range potentials in simulations (SI Fig.~S9). Such self-reinforcing interactions are characteristic of chemotaxis, hence our particles perform ``artificial chemotaxis''. 

Cluster analysis in time provides more quantitative insights into the aggregation behavior (Fig.~\ref{Fig:4}C-E). For each image, we group particles in clusters if their center-to-center distance is smaller than a cutoff distance ($3R$, see Methods). For weak interactions at voltages below 4~\Vpp, all the cluster parameters indicate the system reaches a steady state,   Fig.~\ref{Fig:4}C-E. A finite fraction of single particles remains, Fig.~\ref{Fig:4}C. The mean size of a cluster is finite, as given via the degree of clustering Fig.~\ref{Fig:4}D, $\theta = 1 - 1/\langle n_{\rm clust}\rangle $, where $\langle n_{\rm clust}\rangle$ is the mean cluster size~\cite{mognetti2013living}. The total number of particles in the field of view remains constant within the noise level, Fig.~\ref{Fig:4}E. Cluster properties are rather noisy, indicating that clusters are transiently forming and splitting. We are thus in a dynamic steady state, characteristic of solid-liquid coexistence. In contrast, for high voltages, above 4~\Vpp, the system does not reach a steady state: in particular, the number of particles in the field of view increases with time, characteristic of the collapsing behavior. However, the fraction of single particles rapidly reaches 0, meaning that after an initially short time, coarsening is the dominant growth mechanism. The degree of clustering rapidly reaches 1, characteristic of a solid. Our system thus allows one to easily manipulate various 2D aggregation regimes through field-driven interactions.

\subsection*{Versatility: single-particle beacons and the emergence of non-reciprocity} 
\begin{figure}[htp!]
\includegraphics[trim=0 0 0 0,clip,width=0.45\textwidth]{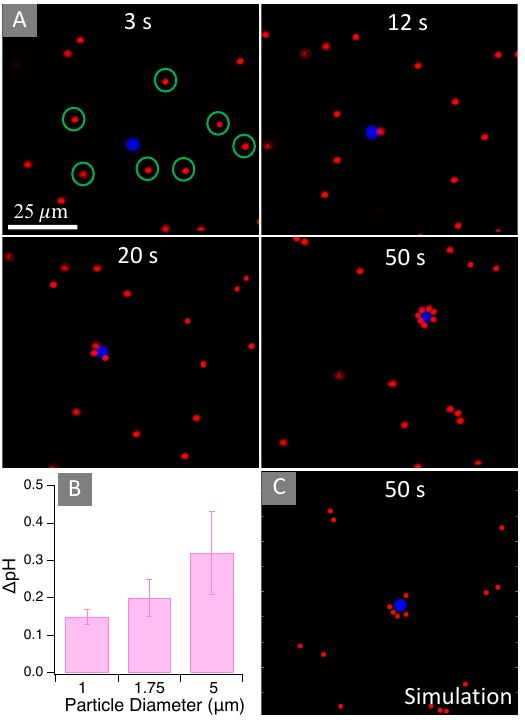}
 \centering
\caption{\label{Fig:5}\footnotesize \textbf{Interparticle potential mediates capture of smaller particles.} (A) Sequence of images for 5 $\mu$m (blue) surrounded by 2 $\mu$m (red) CB-PS particles at the levitated layer under 5 V\TSub{pp} and 100 Hz. A large particle behaves as a beacon for smaller particles, indicated by green circles.  (B) Size dependence of \dpht. Error bars in Panel B represent the standard deviation of values from ten frames and ten different particles, for a total of 100 points. (C) Characteristic result of a Brownian dynamics simulation with interparticle potentials fitted from experimental data and initial configurations from (A, 3s).  
}
\end{figure}

Beyond equal-size aggregation, the versatility of our system is further illustrated by mixing particles of multiple sizes. 
Figure \ref{Fig:5}A shows time-lapse images for 5 $\mu$m (blue) surrounded by 2 $\mu$m (red) CB-PS particles at the levitated layer (see also SI-Movie 4). Initially, small particles are located tens of microns away from a larger blue particle. After about 10~s, small particles begin to migrate towards the large one. Within 50~s, all particles indicated by the green circles eventually become satellites of the large particle. The small particles remain mobile and continue to sample the space adjacent to the large particles through Brownian motion. In the meantime, distant particles sometimes assemble into small transient clusters (as can be seen through a group of 3 at 50~s). By separately tracking the trajectory of a small particle near a large one, it becomes evident that small-large particles experience mutual attraction (SI Fig.~S6). 

To shed further light, we repeat our \dpht~investigation of Fig.~\ref{Fig:3} with particles of different sizes. We find that \dpht~increases with particle diameter (Fig.~\ref{Fig:5}B). Since we have shown \dpht~is a measure of the diffusiophoretic flows surrounding a particle, these measurements suggest the attraction velocities  around a large blue particle are greater than that surrounding a small red particle. The size dependence of $\Delta$pH thus manifests in a non-trivial way on the assembly of particles: the large particles serve as levitating beacons for smaller particles. In terms of interparticle potentials, this would mean that the potential felt by a particle near a large blue particle is greater than that felt near a small red one. As a result, the attraction of red to blue is greater than that from blue to red or red to red, a key feature of non-reciprocal interactions, first demonstrated in active anisotropic colloids that locally consume fuel~\cite{soto2014self}. It is, therefore, surprising to see non-reciprocal interactions in this context -- and it hints at the generality of such a feature for systems with diffusiophoretic flows~\cite{liebchen2021interactions}. 

To confirm the underlying mechanism, we implement Brownian dynamics simulations of particles in 2D with non-reciprocal interactions under the form of standard potentials (see SI Sec.~4). The potentials are calibrated by using experimental data and by making reasonable assumptions when data is scarce. Starting from a similar configuration as in the experiment, we observe a similar beacon-like effect, where small red particles aggregate around the large blue particle over a short time scale of about 50~s (see Fig.~\ref{Fig:4}C and SI-Movie 5). Therefore, using interparticle potentials with minimal features are sufficient to reproduce the essential experimental observations. Electrodiffusiophoresis thus opens new possibilities to manipulate particles and their interactions with electric fields.

\subsection*{Conclusions \& Implications} 
In this article, we have subjected charged particles to complex, non-monotonic, 3D chemical gradients externally generated by electric fields. Under these conditions, driving the electrical double layer far from equilibrium leads to local chemical fields, which in turn drive diffusiophoretic flows and versatile interparticle interactions.  The observed interactions are long-ranged and easily tunable by the external field. Therefore, our findings have expanded the scope of interactions achievable under electric fields beyond dipole-dipole and electrohydrodynamic contributions~\cite{Harraq.2022,Ma:2015fc}. This possibility, which was foreseen in the former Soviet Union by Gamayunov et al.~\cite{Gamayunov.1986}, has rarely been explored. The insights from our work can inform the solution of long-standing problems in colloidal electrokinetics. For example, despite outstanding experimental and theoretical efforts, it has not been possible to explain how electrolyte~\cite{Hoggard:2007gf,Woehl.2014,Wirth:2013dh,Amrei:2018ja} and surface chemistry~\cite{Yang.2019} determine whether particles will separate or aggregate under low frequency AC fields near electrodes. 
Since the diffusiophoretic interactions discussed here are sensitive to surface chemistry and nature of electrolyte, they could play a significant role in inducing aggregation, calling for further theoretical and experimental investigations in that direction.

The adjustable range and magnitude of the interparticle interactions in our setup result in diverse collective dynamics, including states reminiscent of chemotactic collapse and living crystals. In addition, single large particles can behave as beacons, which suggests a strategy to facilitate the migration of analytes towards an interface or to study exotic states in binary mixtures with nonreciprocal interactions~\cite{loos2020irreversibility,osat2023non}. Notice that, unlike common active colloidal systems~\cite{Mu.2022,Fadda.2023,Zheng.2023,theurkauff2012dynamic,Pohl:2014hn,Moran.2017,massana2018active}, particles display complex collective behavior without fuel consumption on their surface, hence our denomination as ``artificial'' chemotaxis. Going forward, these experimental systems offer, on the one hand, an opportunity to create particles with higher complexity by chemically breaking their symmetry since the force is not indifferent to the nature of the surface groups. On the other hand, such 2D, freely levitating systems naturally lend themselves to the study of collective dynamics of 2D colloidal suspensions without the need for a fluid-fluid interface, which can disturb interparticle interactions~\cite{bleibel2014hydrodynamic,bleibel20153d,pelaez2018hydrodynamic}. Diffusiophoretic engineering thus opens numerous opportunities for fundamental studies on colloidal suspensions beyond simple solutes and geometries~\cite{Warren.2020,williams2024colloidal,Raj.2023}.

\section*{Materials and Methods}

The following segments describe the procedures followed for preparation of materials, experimental setup and analysis. 

\subsection*{Model Systems}
The model systems used were fluorescent polystyrene particles (PS) with nominal diameters of 2 $\mu$m (Bangs Lab, FSPP005); fluorescent carboxyl-functionalized polystyrene particles (CB-PS) with nominal diameters of 1 $\mu$m (Bangs Lab, FCGB006), 1.75 $\mu$m (Polysciences, 17686) and 5 $\mu$m (Bangs Lab, FCGB008); and silica particles with nominal diameters of 2.3 $\mu$m (Spherotech, SIP-30-10). For the study of interparticle forces, all particles were dispersed in 100 $\mu$M solutions of SNARF-1 with a concentration of approximately  $5\times10^{-4}\wvp$. Ultrapure deionized water (18 M$\Omega$ cm\textsuperscript{-1}) was used  in all experiments. Measurements of zeta potential ($\zeta$) were performed in a Litesizer 500 (Anton Paar) through electrophoretic light scattering. The particles used for experiments reported in Fig. \ref{Fig:2} were negatively charged, with an average $\zeta$ value of -65.0 $\pm$ 3.1 mV for 1.75 $\mu$m CB-PS, -48.2 $\pm$ 4.4 for 2.0 $\mu$m PS, and -50.2 $\pm$ 5.5 for 2.3 $\mu$m silica particles. For the studies reported in Fig. \ref{Fig:3}E, 5 $\mu$m CB-PS particles were modified by cross-linking poly(ethylene glycol) (PEG) chains of different molecular weights (5, and 10 kDa) to the carboxylate groups (PEG-modified particles, PEG-PS), following the procedure reported in reference \cite{SilveraBatista:2017hn}. Particles became less negatively charged after the attachment of PEG molecules, with the magnitude of change modulated by the molecular weight of PEG. $\zeta$ for the 10 kDa PEG-PS and 5 kDa PEG-PS were -46.2 $\pm$ 2.1 mV and -55.1 $\pm$ 1.9 mV, respectively.

\subsection*{Experimental Setup}
Fig. \ref{Fig:1}A shows a drawing of the experimental setup. The electrochemical cell was built by separating two indium tin oxide (ITO) coated glass slides (Diamond Coatings, 8–10 $\Omega$/cm\textsuperscript{2}) with a dielectric spacer with nominal thickness of 120 $\mu$m (diameter 9 mm, Grace Biolabs, Cat. \# 654002). The ITO slides were positioned such that the ITO coatings were in contact with the suspension. Prior to each experiment, the slides were cleaned by sequentially sonicating them in acetone, isopropanol and DI water for 10 min in each solvent. Then, right before assembling the devices, the slides were treated with UV-ozone (UVO Cleaner Model 30, Jelight) for 10 min. In a typical experiment, approximately 15 $\mu$L of suspension was confined in the electrochemical cell. The AC electric fields were applied using a function generator (Rigol DG1022) with frequency ranging from 100 Hz to 10 kHz, and voltage varying from 1 to 5 Vpp (peak to peak). It is important to mention that we noticed a variability in the electrochemical response of ITO slides from one batch to another. Slides from the same batch behave almost identically. This was observed for slides from Diamond Coatings as well as SPI. 

The particles, fluorescent dye, and electrodes were simultaneously imaged using a Leica SP8 Confocal Laser Scanning Microscope(CLSM). Water (40$\times$, 1.10 NA) and oil (63$\times$, 1.30 NA) immersion objectives were used, while the pinhole was set to 1 Airy unit. The optical properties of particles were selected to avoid significant overlap with the ratiometric dye (SNARF-1). The particles were excited at 405 nm and their emission was collected at wavelengths between 420 and 470 nm. A high-speed resonant scanner (8 kHz) enabled high acquisition rates of up to 28 frames per second at a resolution of 512 $\times$ 512 pixels. Two imaging modes were utilized. XYZT mode was used for observation within a given imaging volume, whereas XYT mode was used to image in $xy$ plane at a given height. To measure the pH around a single particle, the acquisition parameters were optimized to balance acquisition speed with the level of pixel-to-pixel noise. Once a single particle was located, images of approximately $17\times17~\mu\mathrm{m^2}$ in size (with a zoom factor equal to 7), were obtained with $512\times 512$ pixel resolution, and with suitable frame and line averages.   Videos of ten frames were acquired for ten different particles to quantify statistical significance. Further details on imaging can be found in SI Sec. 3. 

\subsection*{Measuring pH during Operation}
Details on the measurement of pH in-operando can be found in our recent publication \cite{Wang.2022p2}. Briefly, pH was mapped by using a ratiometric fluorescent pH indicator, 5-(and-6)-carboxy SNARF-1 (ThermoFisher, C1270). SNARF-1 fluorescent emission presents a shift from yellow-orange to deep red with increasing pH, allowing the pH to be obtained at any point in the experiment through the ratio of dual emissions at two different wavelengths. Using the ratio of dual emission signals minimizes the effect of fluctuations in focus, excitation intensities,  concentration of the dye, and transmittance loss of ITO under electric field, thus proving a more reliable quantification of pH.  
In our experiments, SNARF-1 (100 $\mu$M) was excited at 514 nm, with dual emissions detection at 580 nm and 640 nm. After acquisition of images, pH maps were calculated and plotted using an algorithm developed in Igor Pro, following the formula below:
\begin{equation}
 \mathrm{pH} = \mathrm{pKa}-log\left(\frac{R_{b}-R}{R-R_{a}}\cdot\frac{I_{b,2}}{I_{a,2}}\right) \label{eqn:ratiometric},
\end{equation}

where $R$ represents the ratio of intensities at two detection points, $I_1$ and $I_2$, while $R_b$ and $R_a$ denote the ratios of intensities at the basic ($I_b$) and acidic ($I_a$) end points. The pH at every z-position was calculated by averaging the intensity of the whole frame for each channel. To quantify the pH around particles, the calculations were performed at every pixel of a single frame. To further improve the signal-to-noise ratio, we applied an average convolution filter with a kernel size equal to 5. 

\subsection*{Data analysis}
To obtain particle positions from images, we detect the location of circles encapsulating particles with a specified size and with an intensity threshold (specifically adjusting Matlab's ``imfindcircles''). Particle positions are then obtained from circle centers. Trajectories and positions for the analysis of interparticle interactions in Fig. 1 were performed using the image analysis suite of the LAX S Leica software due to the use of regions of interest to isolate dimers. 

To obtain statistical properties from particle positions we rely on standard definitions and algorithms. The pair distribution function $g(r)$ is defined as follows
\begin{equation}
 g(r)=\frac{\mathcal{A}}{2\pi r N}\sum_{i=1}^N \sum_{j\neq i} \delta (r - r_{ij})  
\label{eq:gr}
\end{equation}
where N is the number of particles considered in an area $\mathcal{A}$ and $\delta$ is the delta function. \eqref{eq:gr} is used to calculate the pair distribution functions reported in Fig.~\ref{Fig:3}. To define clusters, we separate particles in clusters via a distance (in the $xy$ plane) threshold set to $1.5 \times 2 R$ where we recall that $R$ is a particle's radius, with the algorithm of Ref.~\cite{marcon2021distance}. This allows us to report the cluster size with time in Fig.~\ref{Fig:4}.


\subsection*{Simulations}
Interparticle potentials are specified through a Lennard-Jones-like potential~\cite{wang2020lennard}:
\begin{equation}
    \phi(r \leq r_c) =  - \epsilon \alpha(r_c, \sigma) \left(\left( \frac{\sigma}{r}\right)^2 -1  \right) \left(\left( \frac{r_c}{r}\right) -1  \right)
    \label{eq:potentialmain}
\end{equation}
and $\phi(r \geq r_c) = 0$, where $\alpha$ is a constant set such that the minimum of potential energy is $- \epsilon$, $\epsilon$ is the potential depth, $\sigma$ typically corresponds to the location of the potential minimum and $r_c$ to the potential range. $\epsilon, \sigma$ and $r_c$ are parameters of the potential to be specified and that are calibrated on typical experimental data. The functional form of \eqref{eq:potentialmain} was chosen such that (i) the long-range decay of the potential scales like $1/r$ which corresponds to experimental expectations; (ii) one can set the range of the interaction via choosing $r_c$ separately from the value of the location of the potential minimum. Periodic boundary conditions are used, and a standard Euler-Maruyama scheme is used for integration. A further account of simulations is given in the SI Sec.~4.

\section*{Acknowledgements}

We thank Prof. Michael Bevan and Allec Pelliciotti for facilitating the MATLAB codes to calculate order parameters. We also thank Dr. A.S. Dukhin for pointing out important references. We acknowledge support from the National Science Foundation (NSF) under the CAREER award, Grant No. CBET-2239361. Elements of Fig.1A were created using BioRender.com.



%

\end{document}